\documentstyle [12pt] {article}
\begin{document}
%%%%Start of Text%%%%%%%%%%%%%%%%%%%%%%%%%%%%%%%%%%%%%%%%%%%%%%%%%%%%%%%%%%%%
\pagestyle{empty}
\rightline{\vbox{
\halign{&#\hfil\cr
&NUHEP-TH-92-23\cr
&UCD-92-25\cr
&February 1993\cr
&(revised version)\cr}}}
\bigskip
\bigskip
\bigskip
{\Large\bf
	\centerline{Gluon Fragmentation into Heavy Quarkonium}}
\bigskip
\normalsize

\centerline{Eric Braaten}
\centerline{\sl Department of Physics and Astronomy, Northwestern University,
    Evanston, IL 60208}
\bigskip

\centerline{Tzu Chiang Yuan}
\centerline{\sl Davis Institute for High Energy Physics}
\centerline{\sl Department of Physics, University of California,
    Davis, CA  95616}
\bigskip

\begin{abstract}
The dominant production mechanism for heavy quark-antiquark bound states
in very high energy processes is fragmentation, the splitting of a high
energy parton into a quarkonium state and other partons.
We show that the fragmentation functions $D(z,\mu)$ describing
these processes can be calculated using perturbative QCD.
We calculate the fragmentation functions for
a gluon to split into S-wave quarkonium states to leading order in the
QCD coupling constant.  The leading logarithms of $\mu/m_Q$,
where $\mu$ is the factorization scale and
$m_Q$ is the heavy quark mass, are summed up using
Altarelli-Parisi evolution equations.
\end{abstract}

\vfill\eject\pagestyle{plain}\setcounter{page}{1}

Quantitative evidence for quantum chromodynamics (QCD) as the fundamental
field theory describing the strong interactions has come primarily from
high energy processes involving leptons
and the electroweak gauge bosons.
Such processes are simpler than most purely hadronic processes,
because leptons and electroweak gauge bosons do not have strong interactions.
The next simplest particles as far as the strong interactions are concerned
are heavy quarkonia, the bound states of a heavy quark and antiquark.
While not pointlike, the lowest states in the charmonium and bottomonium
systems have typical radii that are significantly smaller than those of
hadrons containing light quarks.  They have simple internal structure,
consisting primarily of a nonrelativistic quark and antiquark only.
The charmonium and bottomonium systems exhibit a rich spectrum of orbital and
angular excitations.  Thus in addition to being simple enough to be used
as probes of the strong interactions, heavy quarkonia are also a
potentially much richer source of information than leptons and electroweak
gauge bosons.

In most previous studies of the production of heavy quarkonia in high
energy processes, it was implicitly assumed that they are produced by
{\it short distance} mechanisms, in which the heavy quark and antiquark
are created with transverse separations of order $1/E$,
where $E$ is the characteristic energy scale of the process.
In this paper, we point out that the
dominant mechanism at very high energies is {\it fragmentation},
the production of a high
energy parton followed by its splitting into the quarkonium state
and other partons.
The $Q {\bar Q}$ pair are created with a separation of order
$1/m_Q$, where $m_Q$ is the mass of the heavy quark $Q$.
The fragmentation mechanism is often of higher order in the QCD coupling
constant $\alpha_s$ than the short distance mechanism, but it is
enhanced by a factor of $(E/m_Q)^2$ and thus dominates at high energies
$E >> m_Q$.  The fragmentation of a parton into a quarkonium state
is described by a fragmentation function $D(z,\mu)$, where $z$ is the
longitudinal momentum fraction of the quarkonium state and $\mu$ is a
factorization scale.  We calculate to leading order in $\alpha_s$
the fragmentation functions $D(z,m_Q)$ for gluons to split into S-wave
quarkonium states at energy scales $\mu$ of order $m_Q$.  The fragmentation
functions at larger scales $\mu$ are then determined by
Altarelli-Parisi evolution equations which sum up the leading
logarithms of $\mu/m_Q$.

One of the quarkonium processes that is important in hadron collider
physics is the production of charmonium at large transverse momentum $p_T$.
A charmonium state can either be produced directly at large $p_T$
or it can be produced indirectly by the decay of a large $p_T$
$B$-meson or a higher charmonium state with large $p_T$.
In previous calculations of the rate for direct production of charmonium
at large $p_T$ \cite{br}, the dominant mechanisms
were assumed to be short distance processes, in which
a collinear $c {\bar c}$ pair in a color-singlet S-wave state is created
with transverse separation on the order of $1/p_T$.
A typical Feynman diagram which contributes to the production of
the $^1S_0$ charmonium state $\eta_c$
at order $\alpha_s^3$ is the diagram for $g g \rightarrow c {\bar c} g$
shown in Figure 1.  The order-$\alpha_s^4$ radiative corrections
to this process include the Feynman diagram for
$g g \rightarrow c {\bar c} g g$ shown in Figure 2.  In most regions
of phase space, the virtual gluons in Figure 2 are off their mass shells
by amounts of order $p_T$, and the contribution from this diagram
is suppressed relative to the diagram in Figure 1 by a power of
the running coupling constant $\alpha_s(p_T)$.  But there is a part of the
phase space in which the virtual gluon attached to the $c {\bar c}$ pair
in Figure 2 is off-shell by an amount of order $m_c$.
The propagator of this virtual gluon enhances the
cross section by  a factor of $p_T^2/m_c^2$.  At large enough $p_T$,
this easily overwhelms the extra power of the coupling constant $\alpha_s$.
The enhancement is due to the fact that the  $c {\bar c}$ pair can be
produced with transverse separation of order $1/m_c$ instead of $1/p_T$.

A more thorough analysis of the amplitude for $g g \rightarrow \eta_c g g$
reveals that the term that is enhanced by $p_T^2/m_c^2$
can be written in a factored form.
The first factor is the amplitude for the production of a virtual gluon
$g^*$ with high-$p_T$ but low invariant mass $q^2$
via the process $g g \rightarrow g g^*$.  In the limit $q^2 << p_T^2$,
it reduces to the on-shell scattering amplitude for $g g \rightarrow g g$.
The second factor is the propagator $1/q^2$ for the virtual gluon.
The third and final factor is the amplitude for the
process $g^* \rightarrow \eta_c g$, in which an off-shell gluon
fragments into an $\eta_c$ and a gluon.
The factoring  of the amplitude allows the fragmentation contribution
to the differential cross section $d\sigma_{\eta_c}(E)$ for
producing an $\eta_c$ with energy $E >> m_c$
to be written in a factorized form:
\begin{equation} {
d\sigma_{\eta_c}(E) \;\approx\; \int_0^1 dz \; d{\widehat \sigma}_g(E/z)
	\; D_{g \rightarrow \eta_c}(z,m_c) \;,
} \label{fac0} \end{equation}
where $d{\widehat \sigma}_g(E)$ is the differential
cross section for producing a real gluon of energy $E$.
All of the dependence on the energy $E$ appears in the subprocess
cross section $d{\widehat \sigma}_g$,
while all the dependence on the quark mass
$m_c$ is in the fragmentation function $D(z, m_c)$.
The variable $z$ is the longitudinal momentum fraction
of the $\eta_c$ relative to the gluon.
The physical interpretation of (\ref{fac0}) is that an $\eta_c$
of energy $E$ can be produced by first producing a gluon of larger
energy $E/z$ which subsequently splits into an $\eta_c$ carrying
a fraction $z$ of the gluon energy.

The generalization of the leading order formula (\ref{fac0})
to all orders in $\alpha_s$ is straightforward.  At higher orders
in $\alpha_s$, the gluon that splits into
a quarkonium state ${\cal O}$ can itself
arise from the splitting of a higher energy parton
into a collinear gluon.  This splitting process gives rise to
logarithms of $E/m_Q$.  In order to maintain the factorization of the
dependences on $E$ and $m_Q$, it is necessary to introduce a factorization
scale $\mu$:  $\log(E/m_Q) = \log(E/\mu) + \log(\mu/m_Q)$.
To all orders in $\alpha_s$,
the fragmentation contribution to the differential cross section
for producing a quarkonium state ${\cal O}$ with energy $E$
can be written in the factorized form
\begin{equation} {
d\sigma_{\cal O}(E) \;=\; \sum_i \int_0^1 dz \; d{\widehat \sigma}_i(E/z,\mu)
	\; D_{i \rightarrow {\cal O}}(z,\mu) \;,
} \label{fac} \end{equation}
where the sum is over all parton types $i$.  The scale $\mu$ is arbitrary,
but large logarithms of $E/\mu$ in the parton cross section
$d{\widehat \sigma}_i$ can be avoided by choosing $\mu$ on the order of $E$.
Large logarithms of $\mu/m_Q$ then necessarily appear in the
fragmentation functions $D_{i \rightarrow {\cal O}}(z,\mu)$,
but they can be summed up by solving the evolution equations \cite{rdf}
\begin{equation} {
\mu {\partial \ \over \partial \mu} D_{i \rightarrow {\cal O}}(z,\mu)
\;=\; \sum_j \int_z^1 {dy \over y} \; P_{i\rightarrow j}(z/y,\mu)
	\; D_{j \rightarrow {\cal O}}(y,\mu) \;,
} \label{evol} \end{equation}
where $P_{i\rightarrow j}(x,\mu)$ is the Altarelli-Parisi
function for the splitting of the parton of type $i$ into a parton of
type $j$ with longitudinal momentum fraction $x$.
For many applications, calculations to leading order in $\alpha_s$
require only the $g \rightarrow g$ splitting function, which is
\begin{equation} {
P_{g \rightarrow g}(x,\mu) \;=\; {3 \alpha_s(\mu) \over \pi}
	\left( {1-x \over x} + {x \over (1-x)_+} + x(1-x)
	\;+\; {33 - 2 n_f \over 36} \delta(1-x) \right) \;,
} \label{split} \end{equation}
where $n_f$ is the number of light quark flavors.
The boundary condition on this evolution equation is the fragmentation function
$D_{i \rightarrow {\cal O}}(z,m_Q)$ at the scale $m_Q$.  It can be calculated
perturbatively as a series in $\alpha_s(m_Q)$.

We proceed to calculate the fragmentation function
$D_{g \rightarrow \eta_c}(z,m_c)$ for a gluon to split into the $^1S_0$
charmonium state $\eta_c$ to leading order in $\alpha_s(m_c)$.
A process (such as $g g \rightarrow g g$) that produces a
real gluon of 4-momentum $q$ has a matrix element of the form
${\cal M}_\alpha \epsilon^\alpha(q)$, where $\epsilon^\alpha(q)$ is the
polarization 4-vector of the on-shell ($q^2 = 0$) gluon.
In the corresponding fragmentation
process (such as Figure 2), a virtual gluon is produced with large
energy $q_0 >> m_c$ but small invariant mass $s = q^2$ of order $m_c^2$,
and it subsequently fragments into an $\eta_c$ and a real gluon.
The fragmentation probability $\int_0^1 dz D(z,m_c)$ is the ratio of the rates
for these two processes.  In Feynman gauge,
the fragmentation term in the matrix element for the $\eta_c$ production
has the form
${\cal M}_\alpha (-i g^{\alpha \beta} / q^2) {\cal A}_\beta$,
where ${\cal M}_\alpha$ is the matrix element for the
production of the virtual gluon and
${\cal A}_\beta$ is the amplitude for $g^* \rightarrow \eta_c g$.
The fragmentation term is distinguished from the
short distance terms in the matrix element by the presence of the
small denominator $q^2$ of order $m_c^2$.  The amplitude ${\cal A}_\beta$
can be written down using  standard Feynman rules for quarkonium
processes \cite{kks}. Multiplying ${\cal A}_\alpha$ by its complex conjugate
and summing over final colors and spins, we get
\begin{eqnarray}
\sum {\cal A}_\alpha {\cal A}_\beta^*
&=& {16 \pi \over 3} \alpha_s^2 {|R(0)|^2 \over 2 m_c }
{1 \over (s - 4 m_c^2)^2} \Bigg( - (s - 4 m_c^2)^2 g_{\alpha \beta}
\nonumber \\
&& \;+\; 2 (s + 4 m_c^2) (p_\alpha q_\beta + q_\alpha p_\beta)
\;-\; 4 s p_\alpha p_\beta \;-\; 16 m_c^2 q_\alpha q_\beta \Bigg) \;,
\label{Asq} \end{eqnarray}
where $p$ is the 4-momentum of the $\eta_c$.
Terms proportional to $q_\alpha$ or $q_\beta$ are gauge artifacts
and can be dropped.
In an appropriate axial gauge, $q_\alpha$ and $q_\beta$
are of order $m_c^2/q_0$ when contracted with the numerator of the
propagator of the virtual gluon.
In covariant gauges, the $q_\alpha$ and $q_\beta$ terms are not suppressed
but are cancelled by other diagrams.
In the $p_\alpha p_\beta$ term, we can set $p = zq + p_\perp$
up to corrections of order $m_c^2/q_0$,
where $z$ is the longitudinal momentum fraction and $p_\perp$ is the
transverse part of the 4-vector $p$.  In a frame where $q = (q_0,0,0,q_3)$,
$z = (p_0 + p_3)/(q _0 + q_3)$ and $p_\perp = (0,p_1,p_2,0)$.
After averaging over the directions of the transverse momentum,
$p_\perp^\alpha p_\perp^\beta$ can be replaced by
$-g^{\alpha \beta} {{\vec p}_\perp}^{\;2}/2$, up to terms that are suppressed
in axial gauge.  The terms in (\ref{Asq}) that
contribute to fragmentation then reduce to
\begin{equation} {
\sum {\cal A}_\alpha {\cal A}_\beta^*
\;=\; {16 \pi \over 3} \alpha_s^2 {|R(0)|^2 \over 2 m_c }
{1 \over (s - 4 m_c^2)^2} \left( (s - 4 m_c^2)^2
\;-\; 2 (1-z) (zs - 4 m_c^2) s \right)
\left( - g_{\alpha \beta} \right) \;.
} \label{Asqfrag} \end{equation}
We have used the conservation of the $q_0 - q_3$ component
of the 4-momentum in the form
$s = ({{\vec p}_\perp}^{\;2} + 4 m_c^2)/z + {{\vec p}_\perp}^{\;2}/(1-z)$.
At this point, it is easy to
calculate the rate for production of $\eta_c g$
in the limit $q_0^2 >> q^2 \sim m_c^2$
and divide it by the rate for production of an on-shell gluon.
The resulting fragmentation probability is
\begin{equation} {
\int_0^1 dz \; D_{g \rightarrow \eta_c}(z)
\;=\; {\alpha_s^2 \over 3 \pi} {|R(0)|^2 \over 2 m_c }
\int_{4 m_c^2}^\infty ds  \int_{4 m_c^2/s}^1 dz \;
{s^2 + 16 m_c^4 - 2 z (s + 4 m_c^2) s + 2 z^2 s^2
	 \over s^2(s - 4 m_c^2)^2} \;,
} \label{Peta} \end{equation}
where $R(0)$ is the nonrelativistic radial wavefunction at the origin
for the S-wave bound state.
We have increased the upper endpoint of the integration over $s$
to infinity, because the resulting error is of order $m_c^2/q_0^2$,
which we have been consistently neglecting.
Interchanging orders of integration, we can read off the
fragmentation function:
\begin{equation} {
D_{g \rightarrow \eta_c}(z,2 m_c)
\;=\; {1 \over 3 \pi} \alpha_s(2 m_c)^2
	{|R(0)|^2 \over M_{\eta_c}^3 } \;
	\Bigg( 3 z - 2 z^2 + 2 (1-z) \log(1-z) \Bigg)\;.
} \label{Deta} \end{equation}
We have set the scale in the fragmentation function and in the running
coupling constant to $\mu = 2 m_c$, which is the minimum value
of the invariant mass $\sqrt{s}$ of the fragmenting gluon.
In the denominator, we have set $2 m_c = M_{\eta_c}$, which
takes into account the correct phase space limitations and is
accurate up to relativistic corrections.
The value of the S-state wavefunction
at the origin $R(0)$ is determined from the $\psi$ electronic width to be
$|R(0)|^2 = (0.8 \; {\rm GeV})^3$.  We use the value
$\alpha_s(2 m_c) = 0.26$ for the strong coupling constant.

Given the initial fragmentation function (\ref{Deta}),
the fragmentation function is determined at larger values of $\mu$
by solving the evolution equation (\ref{evol}) with (\ref{Deta})
as a boundary condition.  The $z$-dependence of
$D_{g \rightarrow \eta_c}(z,\mu)$ at the energy scales $\mu = 2 m_c$
and $\mu = 20 m_c$ is illustrated in Figure 3.
The evolution causes the fragmentation function
to decrease at large $z$ and to diverge at
$z = 0$.  A physical cross section like (\ref{fac}) will still be
well-behaved, because phase space limitations will place an upper
bound on the parton energy $E/z$ which translates into a lower bound on $z$.
It is evident from Figure 3 that taking into account the evolution
of the fragmentation function can significantly increase
the rate for the production process, particularly at small values of $z$.

The fragmentation function for a gluon into $J/\psi$ can be calculated
to leading order in $\alpha_s$ from the Feynman
diagrams for $g^* \rightarrow \psi g g$. The square of the amplitude
$\sum {\cal A}_\alpha {\cal A}_\beta^*$ for this process
can be extracted from a calculation of the matrix element
for $e^+ e^- \rightarrow \psi g g$ \cite{ks}.
The calculation of the fragmentation function is rather involved and we
present only the final result:
\begin{eqnarray}
D_{g \rightarrow \psi}(z, 2 m_c)
\;=\; {5 \over 144 \pi^2} \alpha_s(2m_c)^3 {|R(0)|^2 \over M_\psi^3 }
\int_0^z dr \int_{(r+z^2)/2z}^{(1+r)/2} dy \;
{1 \over (1-y)^2 (y-r)^2 (y^2-r)^2}
\nonumber \\
\sum_{i=0}^2 z^i \left( f_i(r,y) \;+\; g_i(r,y)
	{1+r-2y \over 2 (y-r) \sqrt{y^2-r}}
	\log{y-r + \sqrt{y^2-r} \over y-r - \sqrt{y^2-r}} \right) \;.
\label{Dpsi} \end{eqnarray}
where the integration variables are $r = 4 m_c^2/s$ and  $y = p \cdot q/s$.
The functions $f_i$ and $g_i$ are
\begin{eqnarray}
f_0(r,y) &=& r^2(1+r)(3+12r+13r^2) \;-\; 16r^2(1+r)(1+3r)y
\nonumber \\
&-& 2r(3-9r-21r^2+7r^3)y^2
\;+\; 8r(4+3r+3r^2)y^3 \;-\; 4r(9-3r-4r^2)y^4
\nonumber \\
&-& 16(1+3r+3r^2)y^5 \;+\; 8(6+7r)y^6 \;-\; 32 y^7 \;,
\label{f0} \\
f_1(r,y) &=& -2r(1+5r+19r^2+7r^3)y \;+\; 96r^2(1+r)y^2
\;+\; 8(1-5r-22r^2-2r^3)y^3
\nonumber \\
&+& 16r(7+3r)y^4 \;-\; 8(5+7r)y^5 \;+\; 32y^6 \;,
\label{f1} \\
f_2(r,y) &=& r(1+5r+19r^2+7r^3) \;-\; 48r^2(1+r)y \;-\; 4(1-5r-22r^2-2r^3)y^2
\nonumber \\
&-& 8r(7+3r)y^3 \;+\; 4(5+7r)y^4 \;-\; 16y^5 \;,
\label{f2} \\
g_0(r,y) &=& r^3(1-r)(3+24r+13r^2) \;-\; 4r^3(7-3r-12r^2)y
\;-\; 2r^3(17+22r-7r^2)y^2
\nonumber \\
&+& 4r^2(13+5r-6r^2)y^3 \;-\; 8r(1+2r+5r^2+2r^3)y^4
\;-\; 8r(3-11r-6r^2)y^5
\nonumber \\
&+& 8(1-2r-5r^2)y^6 \;,
\label{g0} \\
g_1(r,y) &=& -2r^2(1+r)(1-r)(1+7r)y \;+\; 8r^2(1+3r)(1-4r)y^2
\nonumber \\
&+& 4r(1+10r+57r^2+4r^3)y^3
\;-\; 8r(1+29r+6r^2)y^4 \;-\; 8(1-8r-5r^2)y^5 ,
\label{g1} \\
g_2(r,y) &=& r^2(1+r)(1-r)(1+7r) \;-\; 4r^2(1+3r)(1-4r)y
\nonumber \\
&-& 2r(1+10r+57r^2+4r^3)y^2
\;+\; 4r(1+29r+6r^2)y^3 \;+\; 4(1-8r-5r^2)y^4 .
\label{g2} \end{eqnarray}
The integrals over $r$ and $y$ in (\ref{Dpsi})
must be evaluated numerically to
obtain the fragmentation function at the energy scale $\mu = 2 m_c$.
At larger values of $\mu$, it is found
by solving the evolution equation (\ref{evol}) with (\ref{Dpsi})
as a boundary condition.  The $z$-dependence of
$D_{g \rightarrow \psi}(z,\mu)$ at the energy scales $\mu = 2 m_c$
and $\mu = 20 m_c$ is illustrated in Figure 4.  The evolution
causes the fragmentation function to decrease at large $z$ and to
diverge at $z = 0$.

An order of magnitude estimate of the gluon fragmentation contribution
to quarkonium production in any high energy process
can be obtained by multiplying the cross section for producing gluons
of energy $E > 2 m_Q$ by the initial fragmentation probability
$\int_0^1 dz D(z,2m_Q)$.  For the $\eta_c$ and $\psi$,
these probabilities are $4.6 \cdot  10^{-5}$ and $2.8 \cdot 10^{-6}$.
Thus, we can expect the asymptotic production rate of $\psi$
to be more than an order of magnitude smaller than $\eta_c$.
The initial fragmentation function for the splitting of
a gluon into the  $^3S_1$ bottomonium state $\Upsilon$
is  also given by (\ref{Dpsi}),
except that the mass $M_\psi$ is replaced by $M_\Upsilon$,
the scale $2 m_c$ is replaced by $2 m_b$, and $R(0)$ is
the appropriate wavefunction at the origin.
The initial fragmentation probability is about $5.3 \cdot 10^{-7}$,
which is smaller than for $\psi$ by about a factor of 5.

In hadron colliders, short distance processes dominate the direct
production of charmonium at small $p_T$, because fragmentation processes
are suppressed by powers of $\alpha_s(m_c)$.
At sufficiently large $p_T$, fragmentation processes must dominate,
because the short distance processes are suppressed by powers of
$m_c^2/p_T^2$. We can make a quantitative estimate of the $p_T$
at which the crossover occurs by comparing the differential cross
sections for short distance process
with the differential cross section for the gluon
scattering process $g g \rightarrow g g$ multiplied by the appropriate
fragmentation probability. For simplicity, we consider
$90^o$ scattering in the $g g$ center of mass frame.
In terms of $p_T$, we have $s = 4 p_T^2$.  In the limit $p_T >> m_c$,
the differential cross section for $g g \rightarrow g \eta_c$ is \cite{br}
$d \sigma/dt = 81 \pi \alpha_s^3 |R(0)|^2/(256 M_{\eta_c} p_T^6)$.
The differential cross section for $g g \rightarrow g g$ is
$d \sigma/dt = 243 \pi \alpha_s^2 / (128 p_T^4)$.  To allow for
fragmentation of either of the two outgoing gluons,  we
multiply by twice the initial fagmentation probability
$\alpha_s^2 |R(0)|^2 / (9 \pi M_{\eta_c}^3)$.  The resulting differential
cross section exceeds that for the short distance process at
$p_T \approx \sqrt{3 \pi/4 \alpha_s} M_{\eta_c} \approx 3 M_{\eta_c}$.
The differential cross section for $g g \rightarrow g \psi$ is \cite{br}
$d \sigma/dt = 5 \pi \alpha_s^3 |R(0)|^2 M_\psi / ( 128 p_T^8 )$
and for this case we estimate the crossover point to be
$p_T \approx \sqrt{1.05/\alpha_s} M_\psi \approx 2 M_\psi$.
Thus fragmentation should dominate over short distance production
at values of $p_T$ that are being measured in present
collider experiments.

The importance of fragmentation for charmonium production in high energy
processes can also be seen from the surprisingly large rate \cite{bck} for
$Z^0 \rightarrow \psi c {\bar c}$, which is two orders of magnitude larger
than that of $Z^0 \rightarrow \psi g g$.  The explanation for this is that
$Z^0 \rightarrow \psi g g$ is a short distance process,
while $Z^0 \rightarrow \psi c {\bar c}$ includes a fragmentation contribution
enhanced by $(M_Z/M_\psi)^2$.  The enhanced contribution arises
{}from the decay $Z^0 \rightarrow c {\bar c}$, followed by the splitting
$c \rightarrow \psi c$ or ${\bar c} \rightarrow \psi {\bar c}$.
With this insight, the lengthy calculation presented in Ref. \cite{bck}
can be reduced to a simple calculation of the fragmentation function
$D_{c \rightarrow \psi}(z,M_Z)$.
This calculation will be presented elsewhere \cite{bcy}.

The probability for a virtual gluon to decay into a $\psi$
was calculated by Hagiwara, Martin and Stirling \cite{hms}
and used to study $J/\psi$ production from gluon jets at the LEP collider.
They did not calculate the fragmentation function
$D_{g \rightarrow \psi}(z,\mu)$ and thus
were unable to sum up large logarithms of $M_Z/m_c$.
Furthermore, their expression for the fragmentation
probability is missing a factor of $1/(16 \pi^2)$, which explains
the surprisingly large rate that they found for this
$\psi$-production mechanism.

A complete calculation of the fragmentation contribution to $\psi$
production in high energy processes must include the production of the P-wave
charmonium states $\chi_{cJ}$, followed by their radiative decays into $\psi$.
In calculating the fragmentation functions for gluons to split into
P-wave charmonium states, there are two distinct contributions that must
be included at leading order in $\alpha_s$.
The P-wave state can arise either from the
production  of a collinear $c {\bar c}$ pair in a color-singlet P-wave state,
or from the production of a collinear $c {\bar c}$ pair in a color-octet
S-wave state \cite{bbly}.  Calculations of the P-wave fragmentation functions
will be presented elsewhere \cite{by}.

We have shown in this letter that the dominant production mechanism
for quarkonium in high energy processes is fragmentation,
the production of a high energy parton followed by its splitting into
the quarkonium state.  We calculated the fragmentation functions
$D(z,\mu)$ for gluons to split into S-wave quarkonium states to leading
order in $\alpha_s$.  The fragmentation functions satisfy
Altarelli-Parisi evolution equations which can be used to sum up
large logarithms of $\mu/m_Q$.  Most previous calculations of
quarkonium production have considered only short-distance production
mechanisms, because the fragmentation mechanism is often of higher
order in $\alpha_s$.  At high energies $E$, the fragmentation mechanism
dominates because it is enhanced by a factor of $(E/m_Q)^2$.
In the case of charmonium production at large $p_T$ in hadron colliders,
we estimated the transverse momentum at which fragmentation begins to
dominate to be less than 10 GeV.
All previous calculations of quarkonium production
{}from this and other high energy processes must therefore be reexamined,
taking into account the possibility of production by fragmentation.

This work was supported in part by the U.S. Department of Energy,
Division of High Energy Physics, under Grant DE-FG02-91-ER40684.
We would like to acknowledge
G.P. Lepage for suggesting the importance of fragmentation
in the production of quarkonium in high energy processes.
We thank M.L. Mangano for pointing out an error in an earlier
version of this paper.
We would also like to acknowledge useful conversations with
G.T. Bodwin, K. Cheung, J. Gunion, and W.J. Stirling.
\vfill\eject

\noindent{\Large\bf Figure Captions}
\begin{enumerate}
\item A Feynman diagram for $g g \rightarrow c {\bar c} g$
	that contributes to $\eta_c$ production at order $\alpha_s^3$.
\item A Feynman diagram for $g g \rightarrow c {\bar c} g g$
	that contributes to $\eta_c$ production at order $\alpha_s^4$.
\item The fragmentation function $D_{g \rightarrow \eta_c}(z,\mu)$
	as a function of $z$ for $\mu = 2 m_c$ (solid line)
	and $\mu = 20 m_c$ (dotted line).
\item The fragmentation function $D_{g \rightarrow \psi}(z,\mu)$
	as a function of $z$ for $\mu = 2 m_c$ (solid line)
	and $\mu = 20 m_c$ (dotted line).
\end{enumerate}
\vfill\eject

\end{document}